# $Cu_2OSeO_3$ Turns Trigonal with Structural Transformation and Implications for Skyrmions

Alla Arakcheeva,[1,*] Priya Ranjan Baral,[1,+] Wen Hua Bi,[1] Christian Jandl,[2] Oleg Janson,[3] Arnaud Magrez[1,*]

1) Ecole Polytechnique Fédérale de Lausanne, SB, IPHYS, Crystal Growth Facility, Lausanne 1015, Switzerland
2) ELDICO Scientific AG, the Electron Diffraction Company, c/o Switzerland Innovation Park Basel Area AG, Hegenheimermattweg 167 A, 4123 Allschwil, Switzerland.
3) Institute for Theoretical Solid State Physics, Leibniz Institute for Solid State and Materials Research Dresden, 01069 Dresden, Germany

+ Present address of P.R. Baral - Department of Applied Physics and Quantum-Phase Electronics Center, The University of Tokyo, Bunkyo-ku, Japan.

*Corresponding authors: alla.arakcheeva@epfl.ch, arnaud.magrez@epfl.ch

**Abstract**
The formation and characteristics of magnetic skyrmions are strongly governed by the symmetry of the underlying crystal structure. In this study, we report the discovery of a new trigonal polymorph of $Cu_2OSeO_3$, observed exclusively in nanoparticles. Electron diffraction and density functional theory calculations confirm its $R3m$ space group, sharing $C_{3v}$ symmetry with Néel-type skyrmion hosts. This polymorph is likely stabilized by surface effects, suggesting that size-induced structural changes may drive a transformation from Bloch-type to Néel-type skyrmions in $Cu_2OSeO_3$. This hypothesis is consistent with prior unexplained observations of Néel-type skyrmions at the surfaces of bulk crystals, which may result from surface-specific structural distortions**.** Overall, these findings provide insights into the interplay between size, structure, and magnetism, opening pathways for controlling skyrmionic properties in nanoscale systems.

**Main**
Symmetry in crystal structures plays a pivotal role in determining emergent phenomena in condensed matter systems, including unique electronic band structures with robust spin-momentum locking **[1–3]**, time-reversal symmetry broken states **[4,5]**, and topological swirling spin textures known as magnetic skyrmions **[6]**. Skyrmions have garnered significant attention for their potential applications in spintronic devices. Their helicity is highly dependent on the symmetry of their host material. **[7]** Known bulk skyrmion hosts include MnSi **[8]**, FeGe **[9]**, $Fe_{1-x}Co_xSi$ **[10]**, and $Cu_2OSeO_3$ **[11,12]**, which crystallize in the cubic $P2_13$ space group, as well as $GaV_4(S/Se)_8$ in $R3m$ **[13,14]** and $VOSe_2O_5$ in $P4cc$ **[15]**. In these systems, the stabilization of the multiple-q skyrmion lattice (SkL) phase originates from the Dzyaloshinskii-

Moriya interaction (DMI) within the helimagnetic ground state, a consequence of the relativistic spin-orbit coupling [16,17]. Notably, skyrmions in $P2_13$ systems are of the Bloch type, whereas in hosts with R3m symmetry, Néel-type skyrmions are observed.

Among these hosts, $Cu_2OSeO_3$ stands out as the first insulating material in which skyrmions were experimentally discovered. Its insulating nature enables electric-field manipulation of the SkL phase [18–22], a property complemented by other phenomena such as the stabilization of an independent SkL phase at low temperature [23,24] and novel magnetic and functional behaviors [25–28]. Under high pressure, $Cu_2OSeO_3$ undergoes a series of structural phase transitions: first to an orthorhombic $P2_12_12_1$ phase, then to a monoclinic $P2_1$ structure, and finally to a triclinic $P_1$ polymorph [29,30]. Remarkably, pressure has been shown to extend the stability range of the SkL phase up to room temperature [30]. These attributes make $Cu_2OSeO_3$ a model system for advancing skyrmion physics. Recent studies have also demonstrated that when skyrmion hosts are confined in nanoparticles, approaching the size of a single skyrmion, the magnetic phase diagram is significantly altered, leading to modifications of the topological spin textures existing in bulk crystals and even lead to the emergence of novel ones [31–34]. However, due to the finite spin-lattice coupling in the aforementioned SkL hosts, it is imperative to discuss the underlying crystal structure, especially while the particle size approaches the diameter of a single isolated sykrmion.

In this work, we present the discovery of a new polymorph of $Cu_2OSeO_3$. Through detailed crystallographic studies and density functional theory calculations, we show that this novel polymorph crystallizes in the trigonal space group *R3m*, belonging to the same $C_{3v}$ point group symmetry than the Néel-type skyrmions host $GaV_4(S/Se)_8$. This structural change suggests that size effects could potentially drive a transformation from Bloch-type to Néel-type skyrmions in $Cu_2OSeO_3$. Our findings offer a possible explanation for the unexpected observations of Néel-type skyrmions at the surfaces of bulk $Cu_2OSeO_3$ crystals. [35]

For this study, bulk and few hundred micron sized single crystals were grown by chemical vapor transport. They were characterized using X-ray diffraction (XRD). Fifty single crystals exhibited a chiral enantiopure structure, with equal distribution between "right-handed" (denoted as **I**) and "left-handed" (denoted as **I'**) enantiomorphs. Both conform to the chiral space group $P2_13$, with structure **I** matching prior reports on $Cu_2OSeO_3$ structure [7,36]. No new polymorph was found in single crystals with size down to few tens of microns. $Cu_2OSeO_3$ nanoparticles were synthesized via a wet chemical process. Their crystal structure was determined using electron diffraction (ED) (**Figure 1a**).

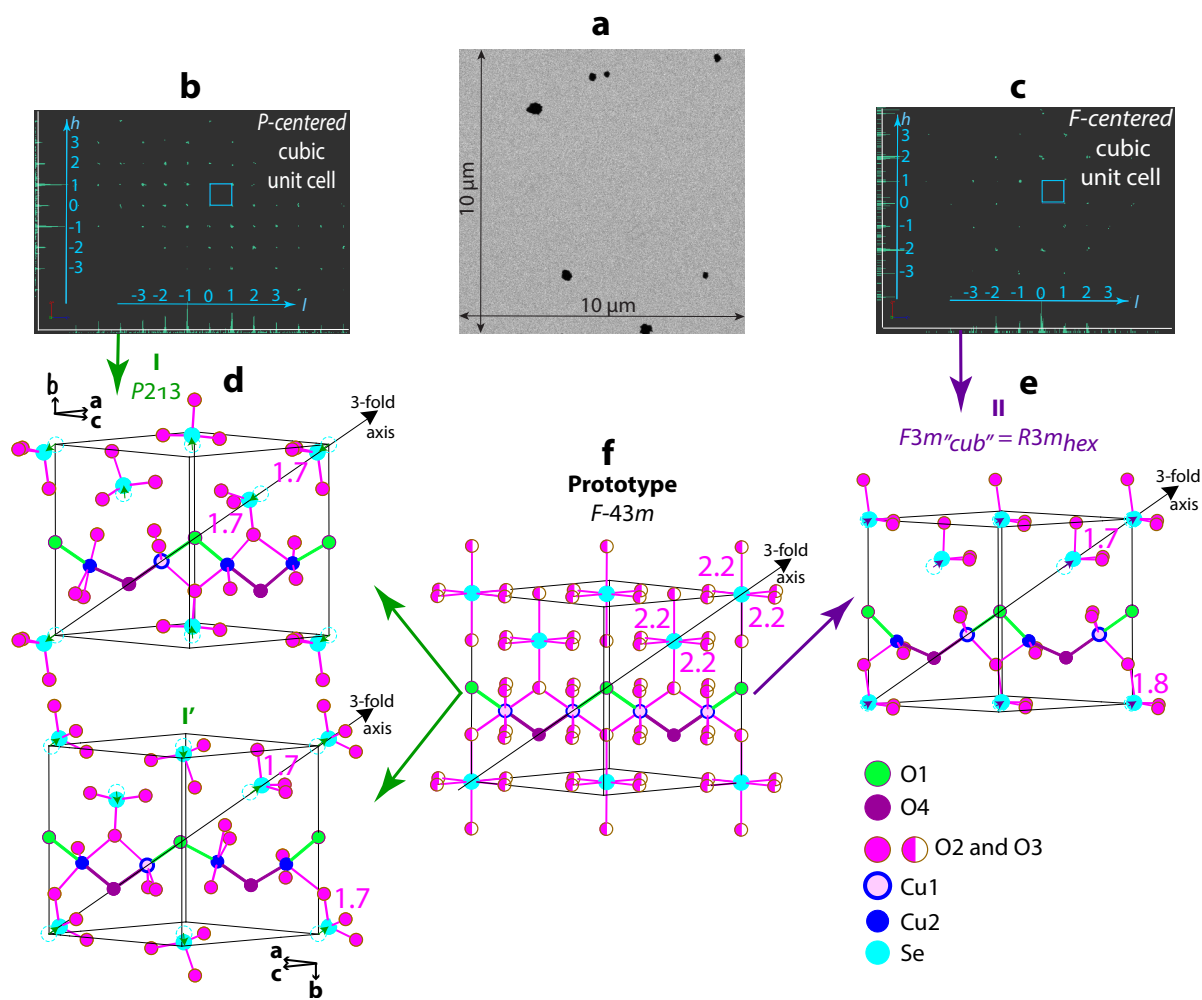

**Figure 1. Structural Interrelations and Symmetry Adaptations in Cu$_2$OSeO$_3$ Crystals and Nanoparticles. (a)** Representative electron microscopy image of the analyzed particles. In (**b**) and (**c**), projection of the electron diffraction patterns along the cubic axis. The absence of extinctions rules in (**b**) points to *P*-centered cubic unit cell characteristic of the crystal structure of both **I** and **I'** enantiomorphs. In (**c**), the reflection extinctions satisfy the condition $h + l = 2n$ revealing an *F*-centered cubic unit cell characteristic for the type-**II** structure. In **d, e** and **f**, representative structural fragments for **I** (**I'**), **II**, and the proposed prototype, showing all independent atoms. All atoms with the exception of O2, O3 (pink) and Cu2 (blue) are located in the 3-fold axes. The substructure Cu$_2$OSe remains identical in all structures. In the prototype (space group *F*-43*m*), the additional oxygen atoms (the half-filled pink circles), required to complete the Cu$_2$OSeO$_3$ structure occupy half-filled positions, resulting in unacceptably long Se–O distances of 2.23 Å. Lowering the symmetry to *P*2$_1$3 for the two enantiomorphic structures **I, I'** or to *R*3*m* (non-standard *F*3*m*) for the structure **II** allows the Se atoms to shift from the (000) position reducing the Se–O distances to approximately 1.70 Å. Concurrently both O2 and O3 occupied sites are increased to full occupancy.

Of the ten analyzed nanoparticles, eight consisted of twins combining both enantiomorphs, typically comprising ~80% of **I** and ~20% of **I'** (**Figure 1b**). Two nanoparticles displayed *F*-centered cubic unit cell (**Figure 1c**), with lattice parameters *a*= 8.893(5)Å (**Table 1**). The corresponding structure (denoted as **II**) was refined in the trigonal space group *R*3*m*

(nonstandard $F3m$) with twinning along the twofold axes (100), (010), and (001) of the cubic basis (**Figure 1e**). Refinement yielded $R_{1obs} = 0.0862$ and $wR_{all} = 0.0656$. To assess the stability of the trigonal structure **II**, density functional theory (DFT) calculations were performed within the generalized gradient approximation (GGA). The results yielded relatively small Hellmann–Feynman forces, suggesting that the structure is close to a local energy minimum. This was further validated by a direct structural optimization, using the experimental unit cell parameters and internal atomic coordinates as input. The total GGA energy of the optimized $R3m$ phase was found to be approximately 0.8 eV/f.u. higher than that of the cubic $P2_13$ phase. However, surface effects, prominent in nanoparticles, cannot be accounted in the DFT calculations but could significantly alter the energy balance and stabilize the $R3m$ phase. Detailed refinement parameters and characteristics of structure **II** are provided in **Tables S1, S2 and S3**. Atomic positions and interatomic parameters of the trigonal phase **II** obtained by refinement of the ED data and by DFT are compared in **Table 2** and **Table S4**, respectively.

All three structures **I**, **I'** and **II** contain a similar substructure unit with $Cu_2OSe$ composition and which has a cubic symmetry with the $F$-43$m$ space group. $Cu_2OSeO_3$ is obtained by adding two O sites with half occupancy (half-filled pink circles in **Figure 1f**). Based on this structural similarity, a prototype model of the ambient-pressure $Cu_2OSeO_3$ structure is proposed. The prototype is constructed from the substructure unit and the two additional O-sites (**Figure 1f** and **Table S5**). However, this leads to unacceptably long Se-O distances of 2.23 Å (**Table S3**). These structure anomalies may be resolved by reducing the structure symmetry in two ways: (i) from $F$-43$m$ to $P2_13$ characteristic of **I** and **I'** (**Figure 1d**) or (ii) from $F$-43$m$ to $R3m_{trigonal}$ = $F3m_{cubic}$ characteristic of **II** (**Figure 1e**). During the refinement, the Se–O distances decrease to an acceptable $1.70 \pm 0.01$ Å in both cases (**Table S3**). We attribute the difference in the crystal structure to the size of the nanoparticles. Indeed, the crystals showing the **I-I'** twinned structure are well-formed nanoparticles as indicated by the rather bright experimental reflections with a low background (**Figure 1b**). The crystals exhibiting the structure **II** are characterized by weaker and split experimental reflections with a much lower intensity (**Figure 1c**), indicating smaller attached nanoparticles. Unlike the pure enantiomorph bulk crystals grown from a single nucleus, multiple nucleation centers form during the synthesis process after selenious acid leaching from $CuSeO_3.2H_2O$ precursor. This leads to nanoparticles with multiple twinning. The smaller particle size observed in the R3m polymorph suggests the existence of a critical size threshold below which the cubic form of $Cu_2OSeO_3$ cannot be stabilized. This hypothesis warrants further investigation.

**Figure 2** illustrates both the similarities and differences among the three structural forms of $Cu_2OSeO_3$. The fundamental building unit consists of two corner-sharing oxygen-centered tetrahedrons, forming structural $[O_2Cu_7]$ dimers **[7]**. Across all structures, the interatomic distances and Cu–O–Cu bond angles remain comparable (**see Table S3** in the Supplementary Information). In the cubic structure, these $[O_2Cu_7]$ dimers exhibit ferrimagnetic ordering, with Cu1 and Cu2 carrying opposing magnetic moments. In all structures, including the prototype (**Figure 2b**), the structural dimers are arranged in hexagonal rings oriented perpendicular to the threefold axis along the four diagonals of the cubic lattice (**Figures 2c-d**). Similar hexagonal arrangements appear in the trigonal lattice along the (001) plane, as well as the (021), (-221), and (2-21) planes (**Figure 2e**). However, the arrangement of the $[O_2Cu_7]$ dimers within these hexagonal rings differs between the cubic and trigonal structures. In the cubic structure, hexagons consist of alternating O1-Cu1-O3 and O3-Cu2-O1 bonds (**Figures 2c-d**), whereas in the trigonal structure, they are built with six O1-Cu2-O4 bonds, with O4-Cu1-O1 acting as bridges along the threefold symmetry axis (**Figure 2e**). The distinct arrangement of dimers in the trigonal structure, as compared to the cubic phase, suggests a different magnetic ordering and hierarchy of energy scales, which may give rise to fundamentally different magnetic structures in the trigonal polymorph.

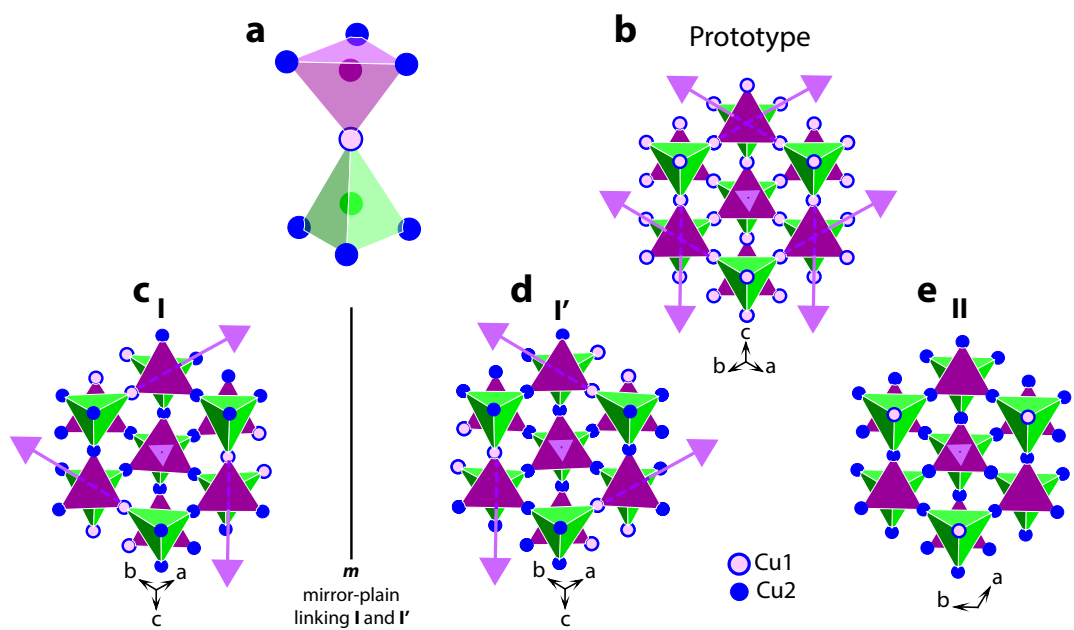

**Figure 2: Structural variations in $Cu_2OSeO_3$.** (**a**) View of the [$O_2Cu_7$] dimer, fundamental structural unit of the $Cu_2OSeO_3$ structures. (**b-e**) Comparison of hexagonal fragments in the prototype structure (**b**), enantiomorphic structures **I** (**c**) and **I'** (**d**), and the trigonal polymorph **II** (**e**). In all panels, copper atoms are shown in blue. Cu1 and Cu2 are respectively located inside and outside the 3-fold axes. The green and purple tetrahedra [$OCu_4$] contain respectively O1 and O4 which are also located inside the 3-fold axes. The light purple arrows and triangles indicate some of the 3-fold axes that distinguish the structures shown. The figure was made using the software Diamond v5.0.2.

In the cation-centered polyhedral representation (**Figure 3**), Cu1 and Cu2 are positioned within a trigonal bipyramid and a tetragonal pyramid, respectively. The Cu2-centered tetragonal pyramids differ between the two structures in terms of Cu–O interatomic distances: in the cubic structure, the apical Cu–O bond is longer than the equatorial ones, whereas in the trigonal structure, all five Cu–O distances are similar. Although the Cu1- and Cu2-containing polyhedra share edges and follow a similar arrangement (**Figure 3a**), the connectivity of Cu2-centered polyhedra varies due to differences in the positioning of $SeO_3$ groups. In the cubic lattice, these tetragonal pyramids contribute to a three-dimensional framework, while in the trigonal structure, they form a flat triangular arrangement (**Figure 3b**). Consequently, the trigonal polymorph of $Cu_2OSeO_3$ exhibits a layered-like structural organization (**Figure 3b**).
This structural distinction may help explain a previously unpredicted observation reported by Zhang et al **[35]** who used resonant elastic X-ray scattering (REXS) at the Cu–$L_2$ absorption edge. At this energy, the X-ray penetration depth is limited to only a few tens of nanometers which is the size range of the $Cu_2OSeO_3$ nanoparticles studied here. The unexpected Néel-type swirls observed at the surface of bulk $Cu_2OSeO_3$ could be attributed to a local symmetry lowering, potentially reflecting the trigonal structure discussed in this work. While our single-crystal XRD results rule out the presence of the $R3m$ polymorph in the bulk, they do not preclude the possibility of this lower-symmetry phase existing at the surface.

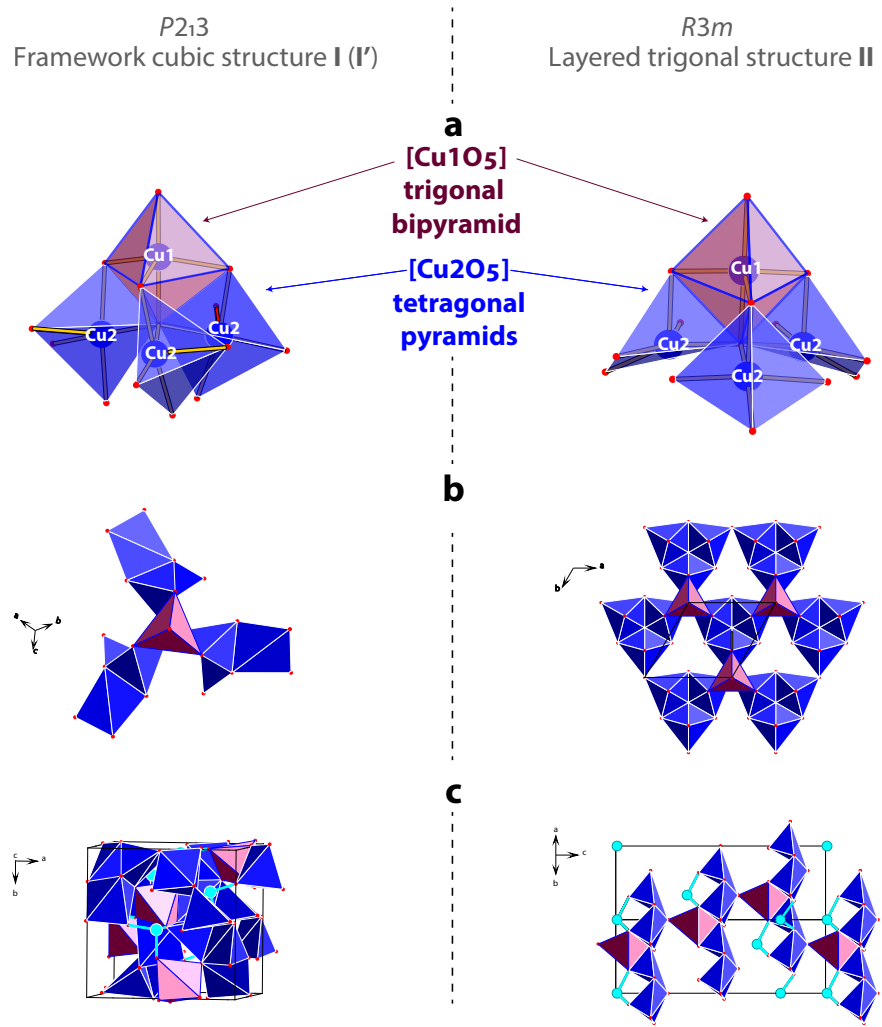

**Figure 3. Cation-Centered Polyhedral Representation of Cubic (I, I') and Trigonal (II) $Cu_2OSeO_3$ Structures**. (**a**) Coordination environments of Cu1 and Cu2 within their respective polyhedra. (**b**) Connectivity of Cu2-centered tetragonal pyramids in the two structures. (**c**) Three-dimensional visualization of the cubic and trigonal frameworks. Selenium atoms are represented as cyan circles. The figure was made using the software Diamond v5.0.2.

In conclusion, this study reveals the discovery of a new polymorph of $Cu_2OSeO_3$, observed exclusively in nanoparticles. Electron diffraction based crystallographic analysis and DFT calculations confirm its $R3m$ space group. While both trigonal and cubic polymorphs share a $[O_2Cu_7]$ dimer-based framework, differences in $SeO_3$ positioning result in distinct connectivity between Cu-centered polyhedra, leading to a layered-like structure in the trigonal phase. The trigonal polymorph exhibits $C_{3v}$ symmetry, like Néel-type skyrmion hosts, suggesting that size effects may drive a transformation from Bloch-type to Néel-type skyrmions in $Cu_2OSeO_3$. This discovery opens several promising directions for future research. First and foremost is the development of synthesis methods or deposition protocols capable of producing phase-pure trigonal nanoparticles or thin films. Such samples would allow precise determination of the stoichiometry of the trigonal phase, either confirming the composition deduced from structural analysis or revealing an off-stoichiometry required to stabilize this polymorph. They would also enable the investigation of skyrmion behavior under $C_{3v}$ symmetry in $Cu_2OSeO_3$. If successful, this would open the door to a range of studies, including the identification of a cycloidal magnetic ground state, the emergence of a field-induced Néel-type skyrmion lattice, potential magnetoelectric coupling, electric-field-driven skyrmion dynamics, and the collective behavior of magnon modes in the microwave regime.

**Table 1. Comparison of Phase Composition in $Cu_2OSeO_3$ Nanoparticles and Single Crystals, Including Previously Published Data**

| Set of tested samples | 10 nanoparticles | | 50 bulk single crystals | | Published data[7,36] |
|---|---|---|---|---|---|
| Phase and its contribution | **I*** and **I'*** as twins; characteristic of 8 nanoparticles | **II**, a new phase; characteristic of 2 nanoparticles | **I***; 50% of the bulk crystals | **I'***; 50% of the bulk crystals | **I*** |
| Space group; system | $P2_13$; cubic | $R3m$ ($F3m$)**; trigonal | $P2_13$; cubic | $P2_13$; cubic | $P2_13$; cubic |
| Unit cell parameters (Å) | $a = 8.890 \div 8.907$ | $a = 6.288 \pm 0.004$ $c = 15.403 \pm 0.010$ ($a^{**}= 8.890 \pm 0.003$) | $a = 8.9134\ (5)$ | $a = 8.9319\ (1)$ | $a = 8.9080\ (5)$ |

*****I** and **I'** are two enantiomorphs of the known cubic phase of $Cu_2OSeO_3$ published in references 7 and 36.
****** The non-standard trigonal space group, $F3m$, and the corresponding pseudo cubic unit cell parameter "$a$" are given for an easy comparison with cubic phase **I**.

**Table 2. $Cu_2OSeO_3$ Atomic Parameters for the Trigonal Phase II ($R3m$ Space Group, $a = 6.284\ (2)$ Å, $c = 15.393\ (3)$ Å) Based on ED Experiments in comparison with DFT calculations**

| | | | Atomic parameters | | | |
|---|---|---|---|---|---|---|
| Atom | Wyckoff position | Source | $x$ | $y$ | $z$ | $U_{iso}$ |
| Cu1 | 3a | ED experiment | 0 | 0 | 3/8 (fixed) | 0.008 (3) |
| | | DFT | | | 3/8 (fixed) | -- |
| Cu2 | 9b | ED experiment | 0.0232 (14) | 0.512 (4) | 0.8527 (9) | 0.042 (3) |
| | | DFT | 0.02185 | 0.51093 | 0.86039 | -- |
| O1 | 3a | ED experiment | 0 | 0 | 0.4991(8) | 0.21 (3) |
| | | DFT | | | 0.50463 | -- |
| O2 | 9b | ED experiment | 0.7454 (5) | 0.8727 (2) | 0.8499 (9) | 0.20 (2) |
| | | DFT | 0.73248 | 0.86627 | 0.84896 | -- |
| O3 | 9b | ED experiment | 0.8195 (10) | 0.639 (2) | 0.3953 (11) | 0.136 (16) |
| | | DFT | 0.81085 | 0.62170 | 0.38562 | -- |
| O4 | 3a | ED experiment | 0 | 0 | 0.2491 (7) | 0.010 (8) |
| | | DFT | | | 0.25206 | -- |
| Se1 | 3a | ED experiment | 0 | 0 | 0.0070 (11) | 0.214 (15) |
| | | DFT | | | 0.00275 | -- |
| Se2 | 3a | ED experiment | 0 | 0 | 0.7863(9) | 0.005 (2) |
| | | DFT | | | 0.78981 | -- |

**Methods**

**Sample preparation**

The synthesis of $Cu_2OSeO_3$ nanoparticles was carried out through a low-temperature reflux process in a round-bottom flask, maintained at a temperature of 70°C. The solution was subjected to continuous stirring throughout the synthesis. The selenite precursor, $CuSeO_3 \cdot 2H_2O$, was synthesized via instantaneous precipitation by mixing highly concentrated solutions of $CuSO_4 \cdot 5H_2O$ and $SeO_2$. During the reflux process, $NH_4OH$ was employed as the base to convert $CuSeO_3 \cdot 2H_2O$ into $Cu_2OSeO_3$. **[29]**

Single crystals of $Cu_2OSeO_3$ were grown using the chemical vapor transport (CVT) method. Crystals ranging from a few hundred microns to several centimeters in size were obtained after two to three weeks, depending on the choice of transport agent. **[7,37–39]** The growth was carried out from a stoichiometric mixture of CuO and $SeO_2$, sealed together with a transport agent in an ampoule measuring 4 cm in diameter and 20 cm in length. To demonstrate that the enantiopure nature of the crystals grown by the CVT is independent of the transport agent, crystals analyzed by XRD were grown using different transport agents: namely $NH_4Cl$, HCl, $Cl_2$ and HBr. The sealed ampoules were placed in a two-zone furnace under a controlled temperature gradient, with the source and the growth zones maintained at 650°C and 590°C, respectively.

**X-ray study**

The XRD experiments were performed on a Rigaku Synergy-I XtaLAB X-ray diffractometer, equipped with Mo micro-focusing source ($\lambda K_\alpha$ = 0.71073 Å) and HyPix-3000 Hybrid Pixel Array detector at room temperature. *Crys-AlisPro* **[40]** and SHELX software **[41]**, were used for the raw experimental data processing and structural refinements, respectively.

**Electron diffraction study**

Samples were deposited as a fine powder on standard TEM grids (amorphous carbon on Cu) and measured on an ELDICO *ED-1* electron diffractometer at room temperature using the software ELDIX (ref. 17). The device uses a $LaB_6$ electron source operating at an acceleration voltage of 160 kV ($\lambda$ = 0.02851 Å; electron exposure – 0.008 $e^-Å^{-2}s^{-1}$;) and a hybrid-pixel direct electron detector (Dectris QUADRO) with pixel size of 75 μm. Imaging of the sample was performed in STEM mode and diffraction was recorded in continuous rotation mode from -60 to +60° with 0.5° rotation per frame and exposure time of 1 sec per frame. A parallel beam of diameter ca. 750 nm was used. The corresponding map resolution covers the range 0.055 – 20.0 Å. The initial frame number was 240 for each data set. The experimental frames were merged to virtual ones for dynamical refinement. Proceeding of the data collections to obtain integrated intensities for ten different sets of reflections were executed with the PETS2 software. **[42]** The structure solutions and structure refinements were performed using Superflip, **[43]** and JANA2020 software, **[44]** correspondingly. Details of the structure refinements and experimental data are given in Table S1 and S2, respectively.

**Equivalent transformation of atomic positions**

To successfully compare representations of previously published structures, and structures shown in figures in the text and in our CIF files, atomic coordinates sometimes need to be transformed using the matrix (0 -1 0 / -1 0 0 / 0 0 -1) with an appropriate shift ( ¼ ¼ ¼ ) as well with the shift of origin by (±0.5 ±0.5 ±0.5), which provides an equivalent transformation in the space group $P2_13$.

**DFT calculations**

For density functional theory (DFT) calculations, we used the generalized gradient approximation (GGA) **[45]** as implemented in the full-potential local-orbital code FPLO version 21. **[46]** The Brillouin zone was sampled with 17x17x17 k-points (969 k-points in the irreducible wedge). Structural optimizations were performed using scalar-relativistic nonmagnetic GGA calculations in the space group R3m (160) without applying the additional local Coulomb repulsion U. Residual Hellmann–Feynman forces were below $10^{-3}$ eV/Å.

**Data availability**

The data that support the findings within this paper are available from the corresponding authors upon reasonable request. The characteristics of structure **II** are available in CSD database with deposition number 2446496. The deposition number of both cubic phases **I** and **I'** is 2408775.

**Acknowledgements**

We thank Lukas Palatinus for his help to use PETS2 program. This work was supported by the Swiss National Science Fundation (SNSF) Sinergia network NanoSkyrmionics (grant No. CRSII5-171003).

**Authors and Affiliations**

A. Arakcheeva, A. Magrez, W.H. Bi and P.R. Baral - Ecole Polytechnique Fédérale de Lausanne, SB, IPHYS, Crystal Growth Facility, Lausanne 1015, Switzerland; Oleg Janson - Institute for Theoretical Solid State Physics, Leibniz Institute for Solid State and Materials Research Dresden, 01069 Dresden, Germany; Christian Jandl – ELDICO Scientific AG, the Electron Diffraction Company, c/o Switzerland Innovation Park Basel Area AG, Hegenheimermattweg 167 A, 4123 Allschwil, Switzerland. Present address of P.R. Baral - Department of Applied Physics and Quantum-Phase Electronics Center, The University of Tokyo, Bunkyo-ku, Japan.

**Contributions**

A.M. initiated and supervised the study. A.A. performed the crystal structure investigations. C.J. performed the ED experiments. W.H.B. and P.R.B. performed the XRD experiments. P.R.B. and A.M. prepared the crystals and nanostructures. O.J. performed DFT calculations. A.A. and A.M. prepared the first draft of the manuscript. All authors read, commented and revised the manuscript.

**Ethics declarations**

Competing interests

The authors declare no competing interests.

# Supplementary information
# Cu$_2$OSeO$_3$ Turns Trigonal: Structural Transformation with Implications for Skyrmions

Alla Arakcheeva*, Priya Ranjan Baral, Wen Hua Bi, Christian Jandl, Oleg Janson, Arnaud Magrez*

*Corresponding authors: alla.arakcheeva@epfl.ch, arnaud.magrez@epfl.ch

**Table S1. Main characteristics of structure refinements for 10 data sets obtained from 10 different nanoparticles in comparison with single crystals**

| Data set | Space group; system | Unit cell parameters (Å) | No. of refl. in refinement: obs.; all | Criteria of obs.: $I/\sigma I$ | R1$_{obs}$; wR$_{obs}$ wR$_{all}$ | Twin element | Twin composition* |
|---|---|---|---|---|---|---|---|
| **Nanoparticles** | | | | | | | |
| 1 (dynamic refinement) | $P2_13$ cubic | $a$ = 8.8946 (7) | 150; 332 | 3 | 0.1198; 0.2274 0.2516 | Mirror plane (-1 1 0) | **I** - 73 (3) %; **I'** - 27 (3) % |
| 2 (dynamic refinement) | $P2_13$ cubic | $a$ = 8.9007 (7) | 680; 853 | 1.5 | 0.1138; 0.1604 0.16 75 | Mirror plane (-1 1 0) | **I** - 86 (1) %; **I'** - 14 (1) % |
| 3 (dynamic refinement) | $P2_13$ cubic | $a$ = 8.9031 (7) | 350; 1667 | 3 | 0.1056; 0.1054 0.1475 | Mirror plane (-1 1 0) | **I** - 85 (1) %; **I'** - 15 (1) % |
| 4 (dynamic refinement) | $P2_13$ cubic | $a$ = 8.901 (7) | 666; 1464 | 1.5 | 0.1501; 0.1203 0.1440 | Mirror plane (-1 1 0) | **I** - 92 (1) %; **I'** - 8 (1) % |
| 5 (kinematic refinement) | $R3m$ trigonal (New phase) | $a$ = 6.2842 (19) $c$ = 15.393 (3) | 246; 452 | 1.5 | 0.0862 0.0550 0.0656 | 2-fold axes along (100), (010) and (001) of a pseudo cubic cell | **II** - 38.8 (15) %; 24.6 (11) %; 36.6 (11) %) |
| 6 (kinematic refinement) | $P2_13$ cubic | $a$ = 8.9070 (8) | 164; 479 | 1.5 | 0.1690; 0.1195 0.1433 | Mirror plane (-1 1 0) | **I** - 78 (2) %: **I'** - 22 (2) % |
| 7 (kinematic refinement) | $R3m$ trigonal (New phase) | $a$ = 6.2923 (8) $c$ = 15.413 (9) | 58; 194 | 1.5 | 0.1741; 0.1546 0.1935 | 2-fold axis along (100) of a pseudo cubic cell | **II** - (47 (5) %; 53 (5) %) |
| 8 (dynamic refinement) | $P2_13$ cubic | $a$ = 8.8900 (5) | 307; 1410 | 3 | 0.0906; 0.0973 0.1388 | Mirror plane (-1 1 0) | **I** - 82 (1) %; **I'** - 18 (1) % |
| 9 (dynamic refinement) | $P2_13$ cubic | $a$ = 8.893 (7) | 571; 974 | 1.5 | 0.1728; 0.1632 0.2749 | Mirror plane (-1 1 0) | **I** - 73 (2) %; **I'** - 27 (2) % |
| 10 (dynamic refinement) | $P2_13$ cubic | $a$ = 8.8961 (5) | 385; 1171 | 1.5 | 0.1210; 0.1211 0.1791 | Mirror plane (-1 1 0) | **I**- 74 (2) %: **I'** - 26 (2) % |
| **Single crystals** | | | | | | | |
| 25 single crystals | $P2_13$ cubic | $a$ = 8.9134 (1) | 1066; 1150 | 2 | 0.0275; 0.0628 0.639 | No twinning | **I** – 100 % |
| 25 single crystals | $P2_13$ cubic | $a$ = 8.9319 (1) | 1093; 1158 | 2 | 0.0228; 0.0530 0.0538 | No twinning | **I'** – 100 % |

*__I__ and __I'__ are two enantiomorphs of the known cubic phase of Cu$_2$OSeO$_3$.[1] __II__ is a newly detected phase in nanoparticles of Cu$_2$OSeO$_3$.

**Table S2. The main characteristics of ED data collection and refinement for two different $Cu_2OSeO_3$ phases found in the nanoparticles.**

|  | Phase II<br>Trigonal phase | Phase I<br>Cubic phase |
|---|---|---|
| **Data collection and processing** | | |
| Voltage (kV) | 160 | 160 |
| Wavelength (Å) | 0.02851 | 0.02851 |
| Electron exposure ($e^-Å^{-2}s^{-1}$) | 0.008 | 0.008 |
| Detector distance (mm) | 578.3 | 578.3 |
| Pixel size (μm) | 75 | 75 |
| Angular range (°) | -60 to +60 | -60 to +60 |
| Rotation per frame (°) | 0.5 | 0.5 |
| Exposure time (s) | 1 | 1 |
| Initial frames (no.) | 240 | 240 |
| Final frames (no.) | 240 | 60** |
| Map resolution range (Å) | 5.1 – 0.77 | 6.3 – 0.77 |
| No. of experimental reflections | 774* | 6855* |
| **Refinement** | | |
| Model resolution range (Å) | 0.055 – 20.0 | 0.055 – 20.0 |
| No. of reflections in refinement | 452* | 1410* |
| Model of refinement | Kinematic | Dynamical |
| Crystallographic system | Trigonal | Cubic |

\* The listed characteristics for phase **I** and **II** correspond to data set **5** and **8** in **Table S1**, respectively.
\*\* Experimental frames were merged to virtual ones for dynamical refinement.

**Table S3. Main characteristics of phase I, I' and II of $Cu_2OSeO_3$ obtained in nanopaticles in comparison with single crystals, with published data and with the prototype structure.**

| Set of tested samples | 10 nanoparticles | | 50 bulk single crystals | | Published data[1] | Prototype** |
|---|---|---|---|---|---|---|
| Phase and its contribution | **I** and **I'** enantiomorphs as twins*; characteristic of 8 nanoparticles among 10 | **II**, a new trigonal phase; characteristic of 2 nanoparticles among 10 | **I**, known cubic phase found in 25 single crystals | **I'**, as enantiomorph of **I** found in 25 single crystals | **I**, known cubic phase | Simulated structure |
| Space group; system | $P2_13$; cubic | $R3m$; trigonal ($F3m$ for the cubic setting) | $P2_13$; cubic | $P2_13$; cubic | $P2_13$; cubic | $F$-$43m$; cubic |
| Unit cell parameters, Å | $a = 8.890 \div 8.907$ | $a = 6.288 \pm 0.004$<br>$c = 15.403 \pm 0.010$<br>($a^* = 8.890 \pm 0.003$) | $a = 8.9134$ (5) | $a = 8.93190$ (10) | $a = 8.9080$ (5) | $a = 8.908$ |
| | Atomic coordinats*** | Atomic coordinats*,*** | Atomic coordinats | | | |
| Se1 ($xxx$) | $x = 0.962 \pm 0.005$ | $x = 0.007 \pm 0.003$ | $x = 0.95905$ (4) | $x = 0.96136$ (4) | $x = 0.9590$ (1) | $x = 0$ |
| Se2 ($xxx$) | $x = 0.712 \pm 0.003$ | $x = 0.786 \pm 0.001$ | $x = 0.71143$ (5) | $x = 0.20905$ (4) | $x = 0.7113$ (1) | $x = 0.75$ |
| Cu1 ($xxx$) | $x = 0.388 \pm 0.004$ | $x = 0.375$ | $x = 0.38623$ (6) | $x = 0.63629$ (4) | $x = 0.3861$ (1) | $x = 0.375$ |
| Cu2 ($xyz$) | $x = 0.137 \pm 0.001$;<br>$y = 0.876 \pm 0.002$;<br>$z = 0.629 \pm 0.002$ | $x = 0.108 \pm 0.002$;<br>$y = 0.840 \pm 0.002$;<br>$z = 0.608 \pm 0.002$ | $x = 0.13385$ (6);<br>$y = 0.87885$ (6);<br>$z = 0.62805$ (6) | $x = 0.13629$ (4);<br>$y = 0.86371$ (4);<br>$z = 0.36373$ (4) | $x = 0.1336$ (1);<br>$y = 0.8787$ (1);<br>$z = 0.6281$ (1) | $x = 0.125$;<br>$y = 0.875$;<br>$z = 0.625$<br>(Cu2 = Cu1) |
| O1 ($xxx$) | $x = 0.268 \pm 0.010$ | $x = 0.249 \pm 0.001$ | $x = 0.2627$ (3) | $x = 0.7607$ (1) | $x = 0.2625$ (3) | $x = 0.25$ |
| O2 ($xyz$) | $x = 0.232 \pm 0.001$;<br>$y = 0.183 \pm 0.003$;<br>$z = -0.033 \pm 0.010$ | $x = 0.286 \pm 0.005$;<br>$y = 0.286 \pm 0.005$;<br>$z = -0.033 \pm 0.005$ | $x = 0.2290$ (3);<br>$y = 0.1871$ (3);<br>$z = -0.0319$(10) | $x = 0.2197$ (3);<br>$y = 0.2327$ (3);<br>$z = 0.0201$ (3) | $x = 0.2291$ (3);<br>$y = 0.1870$ (3);<br>$z = -0.0319$ (3) | $x = 0.75$;<br>$y = 0.75$;<br>$z = 0$ |
| O3 ($xyz$) | $x = 0.268 \pm 0.002$;<br>$y = 0.513 \pm 0.002$;<br>$z = 0.033 \pm 0.005$ | $x = 0.710 \pm 0.005$;<br>$y = 0.481 \pm 0.005$;<br>$z = -0.021 \pm 0.005$ | $x = 0.2699$ (5);<br>$y = 0.5173$ (4);<br>$z = 0.0303$ (7) | $x = 0.2607$ (5);<br>$y = 0.7393$ (4);<br>$z = 0.2393$ (7) | $x = 0.2702$ (3);<br>$y = 0.5176$ (3);<br>$z = 0.0307$ (3) | $x = 0.25$;<br>$y = 0$;<br>$z = 0.5$ |
| O4 ($xxx$) | $x = 0.510 \pm 0.003$ | $x = 0.499 \pm 0.001$ | $x = 0.5106$ (4) | $x = 0.5129$ (4) | $x = 0.5104$ (3) | $x = 0.5$ |
| Selected interatomic distances, Å | | | | | | |
| Se1 – O3 | $1.70 \pm 0.01$ | $1.872 \pm 0.005$ | 1.700 (3) | 1.702 (3) | 1.697 (2) | - $O_{½}$ 2.227 ×6 |
| Se2 – O2 | $1.68 \pm 0.01$ | $1.702 \pm 0.005$ | 1.702 (4) | 1.703 (3) | 1.698 (2) | - $O_{½}$ 2.227 ×6 |
| Cu1 – O | $1.85 \pm 0.03$; | $1.915 \pm 0.005$ | 1.908 (6); | 1.908 (5); | 1.906 (2); | - $O_{½}$ 1.929 ×6; |

| | 1.91 ± 0.03<br>2.11 ± 0.01 ×3 | 1.942 ± 0.005;<br>1.994 ± 0.005 ×3 | 1.921 (5);<br>2.079 (3) ×3 | 1.925 (5);<br>2.086 (3) ×3 | 1.918 (2);<br>2.082 (2) ×3 | - O 1.929 ×2 |
|---|---|---|---|---|---|---|
| Cu2 – O, Å | 1.94 ± 0.01;<br>1.95 ± 0.01;<br>1.99 ± 0.01;<br>2.01 ± 0.01;<br>2.30 ± 0.01 | 1.915 ± 0.005;<br>1.952 ± 0.005;<br>1.970 ± 0.005;<br>2.029 ± 0.005 ×2 | 1.9248 (15);<br>1.970 (3);<br>2.017 (3);<br>2.079 (3);<br>2.284 (3) | 1.928 (1);<br>1.974 (2);<br>1.977 (2);<br>2.022 (3);<br>2.290 (3) | 1.924 (2);<br>1.968 (2);<br>1.968 (2);<br>2.016 (2);<br>2.286 (2) | - O½ 1.929 ×6;<br>- O 1.929 ×2 |
| Selected angles (°) | | | | | | |
| Cu – O – Cu angles in two [OCu$_4$]-tetrahedra | 98 ± 1 ×3,<br>120 ± 1 ×3;<br>105 ± 1 ×3,<br>114 ± 1 ×3 | 97 ± 1 ×3,<br>120 ± 1 ×3;<br>99 ± 1 ×3,<br>117 ± 1 ×3 | 104.72 (16) ×3,<br>116.81 (15) ×3;<br>104.24 (18) ×3,<br>113.77 (13) ×3 | 104.66 (14) ×3,<br>113.82 (11) ×3;<br>101.15 (16) ×3,<br>116.88 (13) ×3 | 104.80 (8) ×3,<br>113.71 (9) ×3;<br>101.39 (9) ×3,<br>116.68 (10) ×3 | 109.47 ×6 |
| O – Se – O angles in two [SeO$_3$]-groups | 100 ± 1 ×3;<br>100 ± 1 ×3 | 90 ± 1 ×3;<br>101 ± 1 ×3 | 99.84 (19) ×3;<br>102.38 (13) ×3 | 99.87 (17) ×3;<br>102.43 (12) ×3 | 99.70 (11) ×3;<br>102.35 (9) ×3 | 90 ×6 |

* The unit cell parameter and atomic coordinates for phase **II** are related to the non-standard pseudo cubic unit cell setting of $F\bar{3}m$ symmetry for ease of comparison. For the standard space group $R\bar{3}m$, the atomic coordinates of structure **II** are listed in **Table 2** of the main text and are available in CIF with the number 2446496

**All atomic parameters used for the prototype structure model are listed also in **Table S5**

*** The "±" symbol reflects the spread of parameter value in all measured nanoparticles

**Table S4. Atomic parameters and interatomic characteristics for the trigonal phase II of Cu$_2$OSeO$_3$ based on ED experiments ($R\bar{3}m$ space group, $a$ = 6.284 (2) Å, $c$ = 15.393 (3) Å) in comparison to DFT calculations**

| Atomic parameters | | | | | | |
|---|---|---|---|---|---|---|
| Atom | Wyckoff position | Source | x | y | z | U$_{iso}$ |
| Cu1 | 3a | ED experiment | 0 | 0 | 3/8 (fixed) | 0.008 (3) |
| | | DFT | | | 3/8 (fixed) | -- |
| Cu2 | 9b | ED experiment | 0.0232 (14) | 0.512 (4) | 0.8527 (9) | 0.042 (3) |
| | | DFT | 0.02185 | 0.51093 | 0.86039 | -- |
| O1 | 3a | ED experiment | 0 | 0 | 0.4991(8) | 0.21 (3) |
| | | DFT | | | 0.50463 | -- |
| O2 | 9b | ED experiment | 0.7454 (5) | 0.8727 (2) | 0.8499 (9) | 0.20 (2) |
| | | DFT | 0.73248 | 0.86627 | 0.84896 | -- |
| O3 | 9b | ED experiment | 0.8195 (10) | 0.639 (2) | 0.3953 (11) | 0.136 (16) |
| | | DFT | 0.81085 | 0.62170 | 0.38562 | -- |
| O4 | 3a | ED experiment | 0 | 0 | 0.2491 (7) | 0.010 (8) |
| | | DFT | | | 0.25206 | -- |
| Se1 | 3a | ED experiment | 0 | 0 | 0.0070 (11) | 0.214 (15) |
| | | DFT | | | 0.00275 | -- |
| Se2 | 3a | ED experiment | 0 | 0 | 0.7863(9) | 0.005 (2) |
| | | DFT | | | 0.78981 | -- |
| Selected interatomic distances, Å | | | | | | |
| Se1 – O3 | | ED experiment | 1.867 (15) ×3 | | | |
| | | DFT | 1.74 (3) ×3 | | | |
| Se2 – O2 | | ED experiment | 1.697 (11) ×3 | | | |
| | | DFT | 1.72 (6) ×3 | | | |
| Cu1 – O | | ED experiment | 1.910 (13); 1.937 (11); 1.989 (12) ×3 | | | |
| | | DFT | 1.8924 (7); 1.9953 (8); 2.0653 (7) ×3 | | | |
| Cu2 – O | | ED experiment | 1.947 (11); 1.965 (8); 2.02 (2) ×2; 1.91 (2) | | | |
| | | DFT | 1.9178 (5); 1.9641 (6); 1.9850 (18) ×2; 2.1809 (8) | | | |
| Selected angles, ° | | | | | | |
| Cu – O – Cu angles in two [OCu$_4$]-tetrahedra | | ED experiment | 119.9 (5) ×3, 97.3 (5) ×3; 117.5 (4) ×3, 99.2 (5) ×3 | | | |
| | | DFT | 119.38 (3) ×3, 94.537 (1) ×3; 117.918 (6) ×3, 99.86 (3) ×3 | | | |
| | | ED experiment | 101.0 (8) ×3; 90.0 (6) ×3 | | | |

| | | | |
|---|---|---|---|
| O – Se – O angles in [SeO$_3$]-groups | DFT | 102.33 (2) ×3; 94.49 (2) ×3 | |

**Table S5. Atomic parameters used to model the structure of Cu$_2$OSeO$_3$ prototype (space group *F*-43*m*, *a* = 8.908 Å).**

| Atom | Wyckoff position | Site symmetry | Occupancy | x | y | z |
|---|---|---|---|---|---|---|
| Cu1 | 16e | .3*m* | 1.0 | 3/8 | 3/8 | 3/8 |
| O4 | 4c | -43*m* | 1.0 | ¼ | ¼ | ¼ |
| O2 | 24f | 2.*mm* | ½ | ¼ | ¼ | 0 |
| O3 | 24f | 2.*mm* | ½ | ¼ | 0 | 0 |
| O1 | 4b | -43*m* | 1.0 | ½ | ½ | ½ |
| Se1 | 4a | -43*m* | 1.0 | 0 | 0 | 0 |
| Se2 | 4d | -43*m* | 1.0 | ¾ | ¾ | ¾ |